# Volumetric Light-field Encryption at the Microscopic Scale

Haoyu Li[1,+], Changliang Guo[1,2,+], Inbarasan Muniraj[2], Bryce C. Schroeder[1,3], John T. Sheridan[2] and Shu Jia[1,3,*]

[1]Department of Biomedical Engineering, Stony Brook University, State University of New York, Stony Brook, New York 11794, USA.

[2]School of Electrical, Electronic Engineering, IoE[2] Lab, The SFI-Strategic Research Cluster in Solar Energy Conversion, College of Engineering and Architecture, University College Dublin, Belfield, Dublin 4, Ireland.

[3]Medical Scientist Training Program, Stony Brook University, State University of New York, Stony Brook, New York 11794, USA.

[*]s.jia@stonybrook.edu

[+]These authors contributed equally to this work.

**ABSTRACT**

We report a light-field based method that allows the optical encryption of three-dimensional (3D) volumetric information at the microscopic scale in a single 2D light-field image. The system consists of a microlens array and an array of random phase/amplitude masks. The method utilizes a wave optics model to account for the dominant diffraction effect at this new scale, and the system point-spread function (PSF) serves as the key for encryption and decryption. We successfully developed and demonstrated a deconvolution algorithm to retrieve both spatially multiplexed discrete data and continuous volumetric data from 2D light-field images. Showing that the method is practical for data transmission and storage, we obtained a faithful reconstruction of the 3D volumetric information from a digital copy of the encrypted light-field image. The method represents a new level of optical encryption, paving the way for broad industrial and biomedical applications in processing and securing 3D data at the microscopic scale.

**INTRODUCTION**

The ever-increasing amount of information that individuals and organizations are storing, processing and analyzing drives the demand for information security technologies. Various such techniques, including steganography, cryptography and digital watermarking, have been used to enhance the security and privacy of data[1–3]. Since the inception of double random phase encoding (DRPE)[4], optical technologies have demonstrated remarkable advantages compared to other encryption methodologies. These advantages include parallel processing, system flexibility, multi-dimensional capabilities, and high encryption density with optical signal processing[5–7]. The rapid development of optical cryptosystems takes advantage of the many degrees of freedom available with both real and phase-space optical parameters, such as amplitude, polarization, wavelength, and phase[8–13]. Several optical encryption techniques employing variations of the classical Fourier transform based DRPE system have been further developed, including Fresnel transform (FST), Fractional Fourier transform (FRT), Hartley Transform (HT), Gyrator Transform (GT), and Linear Canonical Transform (LCT)[14–19]. The quantum nature of light has also been explored as a security key in quantum communications[20,21].

Recently, various approaches have been reported for the optical encryption of 3D objects at the macroscopic scale (millimeters, centimeters to meters)[22–31]. However, the existing methods are ineffective at the microscopic scale, despite the ever-growing demand in this regime. This challenge is mainly caused by two factors. First, in macroscopic scenes, the light-field information can be effectively analyzed using geometrical optics, but because diffraction now plays a crucial role, a wave optics model must be considered in the recording and processing of the data at the microscopic level. Second, in macroscopic scenes, optical encryption is primarily restricted to the diffusely reflecting surface of objects due to scattering, which, however, no longer dominates at the microscopic level, leading to most objects being transparent. These factors demand distinct image formation, encryption and processing algorithms for the largely unexplored microscopic scale. An encryption method to address these challenges is highly desired and expected to lay the foundation for many new applications.

Here we report volumetric light-field encryption, a method that allows 3D data at the microscopic scale to be processed and secured as a single 2D image, which can be used to faithfully reconstruct the volumetric information. The method is based on developments in light-field imaging, which utilizes microlens arrays to simultaneously capture both the spatial and angular information of light, allowing the computational synthesis of 3D focal stacks across the entire volume of a 3D object. In the early demonstrations of light-field imaging systems, the use of ray-optics models for reconstruction inevitably compromised the lateral and angular (i.e. axial) resolution[32–35]. Recently, wave-optics models have been developed for light-field reconstruction, which mitigates the resolution loss through PSF deconvolution, thereby significantly advancing high-resolution 3D light-field imaging[36–38]. The algorithm reported in this work integrates the wave-optics models of both a microlens array and an array of random phase masks. We demonstrated that such an encryption system can faithfully recover both multiplexed discrete data and continuous volumetric data. Furthermore, we showed that volumetric information can be accurately recovered from a series of binary images digitally converted from a 2D light-field image, which will potentially be useful for the storage and transmission of large amounts of volumetric data. We anticipate that this light-field method will advance the development of optical encryption methodologies for securing, storing and transmitting volumetric data and have new applications in industry and the life sciences.

**RESULTS**

**Experimental and Computational Encryption Model**

As shown in Figure 1a, the light-field based encryption system is established by placing a microlens array at the intermediate image plane of a wide-field microscope[34]. The microlens array is chosen to match the numerical aperture and magnification of the objective lens. This consideration allows the images formed by each individual lenslet to tile with each other without overlapping or leaving gaps on its back focal plane, where the camera sensor is placed. Each lenslet generates a perspective-view of the object, which propagates angular information onto different regions of the camera. Next, an array of random phase masks, each identical in size to a lenslet, is attached to the microlens array. Random phase masks are typically utilized in optical encryption as a pivotal data security feature. This random modulation of the light field can be experimentally realized using a phase plate or a spatial light modulator, located on a plane that is conjugated to the microlens array. With this optical design, both the spatial intensity and the angular (i.e. axial) information of the volumetric 3D data are encrypted after passing through the optical components. It is then recorded as a 2D light-field image on the camera.

To decrypt the data, we thus need to take into account the light-field information originating from both the microlens array and the random phase masks. For the microlens array, a traditional image retrieval method would generate a low-resolution, perspective-view of the object by tiling those pixels that correspond to the same incidence angle, and a conventional 2D image by summing the pixels behind each lenslet. However, in this case, the axial resolution compromises the lateral resolution, which is substantially limited by the lenslet size. We utilized a wave-optics model that overcomes this resolution limitation by considering the diffraction of light (i.e. the PSF) in the formation of the image through the microlens array and the random phase masks. It should be noted that though there are various types of random phase masks which could be used for the method, using the form of an array of masks, each matching a lenslet, ensures that the known phase information of the masks can be readily integrated into the same wave-optics model as is used for the microlens array.

This integrated model also considers any cross talk generated by the random masks. Hence, the universal PSF, composed of both optical parts in the light-field system, becomes the key by which the original volumetric information is retrieved (see Methods for more details).

Mathematically, the decryption is an inverse problem to recover the radiant intensity at each point in the 3D volume, denoted with **g**, using the recorded camera image **O**. As shown in Figure 1b, the two pieces of information satisfy the relationship $\boldsymbol{O} = H\boldsymbol{g}$, where the measurement matrix $H$ is determined by the PSF, and its elements $h_{kj}$ represent the projection of the light arriving at pixel $\boldsymbol{O}(j)$ on the camera from the $k^{th}$ calculated volume g($k$) in the object space. To obtain **g**, the inverse problem thus becomes:

$$\boldsymbol{g}^{(k+1)} = \mathrm{diag}[\mathrm{diag}(H^T H \boldsymbol{g}^{(k)})^{-1}(H^T \boldsymbol{O})]\boldsymbol{g}^{(k)}. \qquad (1)$$

The operator $\mathrm{diag}()$ diagonalizes a matrix. This expression is a modified deconvolution algorithm[36,37,39] based on the Richardson-Lucy iteration scheme[40].

**3D Data Encryption and Decryption**

To validate the light-field encryption model, we created a volumetric 32-µm x 32-µm x 50-µm object space that consists of volumetric images of the letters 'S', 'B' and 'U', with the same 2-µm thickness, located at -60 µm, -34 µm and -10 µm in the axial dimension, respectively, as shown in Figure 2a. The imaging and encryption system is primarily composed of a 20x, 0.5NA objective, a microlens array with a 150-µm pitch and 3-mm focal length, located at the intermediate image plane, an array of random phase masks (pixel size = 10 µm) with the same pitch, and a CCD camera with a 10-µm pixel size. An emission wavelength of 532 nm from the objects is considered. The 3D image data was encrypted and captured by the camera as a 2D light-field image, as shown in Figure 2b. It should be noted that even without the random phase masks, the microlens array redistributes and projects the 3D image-forming light, protecting the original information from visual scrutiny and thus providing a basic level of data security. Next, the encrypted image was analyzed using the wave-optics based deconvolution algorithm, where the object space was sectioned into 128 x 128 x 26 volumes with a lateral (in x and y) interval of 0.25 µm and an axial (in z) interval of 2 µm. These interval values were empirically chosen to work with the resolution of the objective and the computational capacity available to execute the algorithm. Figure 2c shows the successful restoration of the volumetric images using the correct decryption key, i.e. the correct PSF of the system. We also examined the occlusion attack, where 25% of the image area was blocked (Figure 2d). Figure 2e shows that the method can still recover the volumetric objects ('U', 'B', 'S') with moderately increased noise level. It is also seen that when the similarly occluded area was further increased (e.g. to 37.5%), the 3D object cannot be reconstructed faithfully. In addition, when the values of the elements in such a PSF key are randomly modulated by 5%, the reconstruction becomes noticeably degraded and inaccurate (Figure 2f). This demonstrates that volumetric 3D data can be encrypted into a single 2D light-field image, and that by using the correct PSF key, the original information can be retrieved with high accuracy and sensitivity.

**Multiplexed Optical Processing**

The light-field method allows the encryption of information in an additional axial dimension without increasing the size and complexity of the recorded data. This capability not only increases the information density, allowing volumetric data to be multiplexed and encrypted in a 2D image, but also suggests a means for the direct encryption of a continuous 3D object (as opposed to the discrete volumetric objects in Figure 2). In this Section, to further test the multiplexing capability of the light-field encryption system, i.e. the robustness against crosstalk effects, we positioned three 2-µm thick objects, 'deer', 'dog' and 'bird', at distances of -60 µm, -34 µm and -10 µm in the object space, respectively. In addition, a 3D object, 'bunny', with the size of 30 µm × 30 µm × 50 µm, was placed in the same volume (Figure 3a). In this case, the four objects in the volume were multiplexed in both the lateral and axial dimensions, and encrypted and recorded as a light-field image in Figure 3b. The same 3D deconvolution algorithm was utilized for decryption. In Figure 3c-g, we show the successful recovery of all four volumetric objects. The information has mostly been recovered and the crosstalk between the multiplexed objects is minimal. By increasing the number of iterations of deconvolution and using proper filtering thresholds, we found that the reconstructed objects could be further refined. In order to quantitatively analyze the performance of the decryption method, normalized correlations between the initial objects of 'deer', 'dog' and 'bird', and their respective decrypted images were performed (Figure 3h-j). The correlation peak value *C* in these cases were measured to be >0.7-0.8, indicating a high similarity between the original and reconstructed data. In brief, the results demonstrate that both discrete and continuous spatially-multiplexed volumetric data can be reliably encrypted and decrypted using the light-field method. It also implies that high parallelization can be achieved using the light-field approach as the 2D wide-field image allows various types of data to be parallelized in all three dimensions and processed all at once through the algorithm using the same PSF key.

**Digitization for Binary Storage and Read-out**

For data processing, storage and transmission, digitization allows the data to be operated on with high efficiency and minimal degradation. In this work, the pixel values of the light-field image can be normalized to $2^N - 1$ and rounded to the closest integer values *P* in a *N*-level binary representation, expressed as $P = a_0 \cdot 2^0 + a_1 \cdot 2^1 + a_2 \cdot 2^2 + \cdots + a_{N-1} \cdot 2^{N-1}$, where $a_i = 0 \; or \; 1 \; (i = 1, \cdots, N-1)$ is the pixel-value composite coefficient. The digitization of the light-field image thus refers to the processing of light-field data into *N* binary (i.e. pixel value at 0 or 1, or black or white) images, where image *i* (*i* = 0 ⋯ *N*-1) represents the binary composites of the pixel values of the light-field image at $2^i$. In Figure 4a, the pixel values of the 2D encrypted light-field image were converted into a *N* = 2 level representation (i.e. pixel values were first normalized to $2^N - 1 = 3$, rounded to the closest integer values, and expressed in binary). Hence, two binary images (Figure 4b) were formed containing the pixel-value composites at $2^0 = 1$ and $2^1 = 2$, respectively. For example, a normalized and rounded pixel value 3 can be written into $a_0 \cdot 2^0 + a_1 \cdot 2^1$ ($a_{0,1} = 1$), so the corresponding pixels in the two binary images will both be given the values of 1. As shown in Figure 4c, despite the binarization, decryption using the 2-level digitized images largely reveals the original volumetric information. With an additional binary level ($N$ = 3), the retrieval of the 3D data compares favorably to the reconstruction using the full light-field information as in Figure 2c (Figure 4d-f). Such a digitization procedure transforms a 2D image of full dynamic range into a discrete set of binary images, allowing a substantial reduction in data storage requirements without compromising the quality of retrieval. The conversion of the light-field image into binary images also allows the original data to be partially accessed by multiple users, suggesting another level of utility and security. Moreover, the practicality of digitizing these light-field images allows more convenient and versatile access to the volumetric data. For example, the digitized binary images can be scanned and processed in a similar way to a quick-response (QR) code by a smartphone or tablet device[41]. Digitization also provides a way for transmitting light-field information through digital systems, such as optical fibers. We anticipate more powerful and diverse ways for encrypting, storing and transmitting 3D volumetric data can be achieved with advanced data processing strategies for light-field images.

**DISCUSSION**

In summary, we have developed a new light-field encryption method and demonstrated the encryption and decryption of high-content, microscopic 3D information based on a single 2D light-field image. The wave-optics based PSF key, consisting of the microlens array and the array of random phase masks, allows the retrieval of data with high resolution, accuracy and sensitivity. The method represents a new advance in the optical encryption of 3D image data in the microscopic regime. It should be mentioned that, instead of using phase masks, using amplitude masks can further strengthen data security, completely preventing any visual scrutiny (as some shape of the objects can still be discerned through individual lenslets in **Figures 2 and 3**). A similar light-field encryption system that employs random amplitude masks has been demonstrated in **Supplementary Figures 1 and 2**.

As a new encryption platform in the microscopic regime, we envision broad applications for the method, such as high-density data security, watermarking, 3D scanning, identity authentication, etc[42–44]. Given its high-resolution and volumetric features, the method will be especially useful in the processing of protected 3D biomedical data[45,46]. In addition, the current algorithm is readily compatible with non-binary data (**Supplementary Figure 3**) and fluorescent (or histologic) imaging. It should be noted that the axial dimension in the volumetric data can also be considered as a dimension for time, so that the method is readily adaptable to the encryption of dynamic 2D image data. Another worthwhile future development of the method would be to use different random phase/amplitude masks for each lenslet, a measure which would increase security by expanding the size of the key space. The method could also be enhanced to take advantage of other optical degrees of freedom, such as wavelength, fluorescent life-time, or polarization[47–50], providing further applications of this volumetric multiplexing encryption method. Beyond image data, other types of information, either intrinsically 3D data or stacks of 2D data, can be digitally processed in a similar manner, with the principle of the PSF key being readily extended to other formats.

## METHODS

### Model of Light-field Propagation and Image Formation

When the 3D volume in the object domain is projected into the imaging space, the wavefront at the intermediate image plane, in the case of high numerical aperture microscopy, is predicted by the Debye theory as[51]:

$$U_i(\mathbf{x}, \mathbf{p}) = \frac{M}{f_{obj}^2 \lambda^2} \exp\left[-\frac{iu}{4\sin^2(\alpha/2)}\right] \times \int_0^\alpha P(\theta) \exp\left[\frac{iu\sin^2(\theta/2)}{2\sin^2(\alpha/2)}\right] J_0\left[\frac{\sin(\theta)}{\sin(\alpha)} v\right] \sin(\theta) d\theta, \quad (2)$$

where $f_{obj}$ is the focal length of the objective, and $J_0$ is the zeroth order Bessel function of the first kind. The variables $v$ and $u$ represent normalized radial and axial optical coordinates; the two variables are defined by $v = k[(x_1 - p_1)^2 + (x_2 - p_2)^2]^{1/2} \sin(\alpha)$ and $u = 4kp_3 \sin^2(\alpha/2)$. $\mathbf{p} = (p_1, p_2, p_3)$ is the position for a point source in a volume, and $\mathbf{x} = (x_1, x_2) \in R^2$ represents the coordinates on the sensor plane. $M$ is the magnification of the objective. The half-angle of the numerical aperture $\alpha = \sin^{-1}(NA/n)$ and the wave number $k = 2\pi n/\lambda$ are calculated using the wave length $\lambda$ and the index of refraction $n$ of the sample. For Abbe-sine corrected objectives, the apodization function of the microscope $P(\theta) = \cos(\theta)^{1/2}$ in this case.

The aperture of a lenslet can be described as an amplitude mask $\text{rect}(\mathbf{x}/d)$, combined with a phase mask $\exp\left(\frac{-ik}{2f_{\mu lens}} \|\mathbf{x}\|_2^2\right)$. The modulation induced by a lenslet is then described as

$$\phi(\mathbf{x}) = \text{rect}(\mathbf{x}/d) \exp\left(\frac{-ik}{2f_{\mu lens}} \|\mathbf{x}\|_2^2\right), \quad (3)$$

where $f_{\mu lens}$ is the focal length, and $d$ is the pitch and diameter of a single lenslet. The modulation of the entire lens array, composed of many identical lenslets, can be described by the convolution of $\phi(\mathbf{x})$ with a 2D comb function $\text{comb}(\mathbf{x}/d)$, i.e. $\Phi(\mathbf{x}) = \phi(\mathbf{x}) \otimes \text{comb}(\mathbf{x}/d)$.

The array of random phase masks is positioned immediately after the microlens array, identical in size to each lenslet. A random phase mask behind each lenslet is described as

$$r(\mathbf{x}) = \text{rect}(\mathbf{x}/d) \exp(-i\pi \boldsymbol{\beta}), \quad (4)$$

where $\beta$ is generated by a random number generator, taking values in the range of [-0.5, 0.5]. Similar to the microlens array, the phase modulation of the random phase masks is described as the convolution of $r(\mathbf{x})$ with the same 2D comb function $\text{comb}(\mathbf{x}/d)$, i.e. $R(\mathbf{x}) = r(\mathbf{x}) \otimes \text{comb}(\mathbf{x}/d)$.

In the system shown in Figure 1a, the microlens array and the phase masks are positioned at the intermediate image plane, and the CCD sensor is positioned at a distance of $f_{\mu lens}$ from the microlens array. The light propagation from the microlens array and the array of random phase masks to the CCD sensor can be modelled using the Fresnel approximation over a distance of $f_{\mu lens}$. The final complex PSF, which is computed using the Fourier transform operator $\mathcal{F}\{\}$ and inverse Fourier transform $\mathcal{F}^{-1}\{\}$, is described as

$$h(\mathbf{x}, \mathbf{p}) = \mathcal{F}^{-1}\left\{\mathcal{F}[U_i(\mathbf{x}, \mathbf{p})\Phi(\mathbf{x})R(\mathbf{x})] \times \exp\left[i2\pi\lambda f_{\mu lens}\sqrt{1 - (f_x^2 + f_y^2)}\right]\right\}, \quad (5)$$

where the exponential term is the transfer function of the Fresnel diffraction integral, and $f_x$ and $f_y$ are the $x$ and $y$ spatial frequencies in the sensor plane. Additional discussions about the Fresnel diffraction model can be found in **Supplementary Note**. The final encrypted intensity image $O(\mathbf{x})$ at the CCD plane is described by

$$O(\mathbf{x}) = \int |h(\mathbf{x}, \mathbf{p})|^2 g(\mathbf{p}) d\mathbf{p}. \quad (6)$$

where $\mathbf{p} \in R^3$ is the position in a volume containing isotropic emitters whose combined intensities are distributed according to $g(\mathbf{p})$. In the discrete model of the complex PSF, $h(\mathbf{x}, \mathbf{p})$ is represented by the measurement matrix $H$ which elements $h_{kj}$ represent the projection of the light arriving at pixel $O(j)$ on the camera from the $k$th calculated volume $g(k)$ in the object space, as shown in Figure 1b.

**ACKNOWLEDGMENTS**

This work was supported in part by Stony Brook University.

**AUTHOR CONTRIBUTIONS**

H.L. and C.G. conceived the idea and performed the research. B.C.S. and I.M. helped with the development of the project. J.T.S. provided critical advice and guidance. S.J. supervised the project. All authors contributed to the discussions and participated in the writing of the manuscript.

**ADDITIONAL INFORMATION**

Competing financial interests: The authors declare no competing financial interests.

**FIGURES**

**Figure 1.** (a) Optical setup of the light-field encryption system. The object is imaged by the objective lens, and an intermediate image is formed by the tube lens. The microlens array is located at the intermediate image plane of the tube lens. An array of random phase masks (RPMs) is located next to the microlens array. A camera sensor is placed at a distance of $f_{\mu lens}$ after the microlens array and RPM. (b) Discrete PSF matrix for light-field encryption and decryption. ***g*** describes the 3D volumetric data of the object. ***O*** describes the light-field image on the camera. The measurement matrix *H* describes the relationship $\boldsymbol{O} = H\boldsymbol{g}$, and its elements $h_{kj}$ represent the projection of the light arriving at pixel O(*j*) from the *k*[th] calculated volume g(*k*).

**Figure 2.** Volumetric light-field encryption and decryption. (a) The three volumetric objects of the letters 'S', 'B' and 'U', located at -60 μm, -34 μm, and -10 μm along the axial (*z*) dimension, respectively. (b) The corresponding encrypted 128-pixel x 128-pixel 2D light-field image. (c) Decrypted 3D volumetric information using the correct PSF key. (d) 25% area of the encrypted image in (b) was occluded. (e) The corresponding decrypted 3D volumetric information from the occluded 2D light-field image in (d). (f) Decrypted 3D volumetric information using an incorrect PSF key (5% random errors). Insets in (a), (c), (e) and (f) show the cross-sections of the respective data at -60 μm, -34 μm, and -10 μm along the axial (*z*) dimension. Scale bar in (a), 4 μm.

**Figure 3.** Optical encryption of multiplexed volumetric data. (a) 3D volumetric objects containing discrete and continuous volumetric data. (b) The encrypted 256-pixel x 256-pixel 2D light-field image. (c) Decryption and reconstruction of 3D volumetric information. (d) Zoom-in of the continuous 3D object ('bunny') in (c). (e-g) Lateral cross-sections (in the x-y plane) of the reconstructed volumetric images ('deer', 'dog', 'bird') at -60 μm, -34 μm and -10 μm, respectively. (h-j) Normalized correlations between the original volumetric data and the retrieved data in (e-g), respectively. The correlation peak value C for each of the objects are 'deer', $C$>0.8; 'dog', $C$>0.7; and 'bird', $C$>0.8. Scale bar in (g), 4 μm.

**Figure 4.** Digitization for binary storage and read-out. (a) $N$ = 2 level digitization of the encrypted 2D light-field image in Figure 2b. Pixel values of the light-field image were normalized to $2^N - 1 = 3$ and rounded to the closest integer values $P$, which can be expressed as $P = a_0 \cdot 2^0 + a_1 \cdot 2^1$, where $a_i = 0 \; or \; 1 \; (i = 0, 1)$ is the pixel-value composite coefficient. (b) The respective images that represent the binary composites of the pixel values of the light-field image at $2^0 = 1$ and $2^1 = 2$, respectively. (c) Decrypted volumetric data using digitized images in (b), revealing the original volumetric information. (d) $N$ = 3 level digitization of the same 2D light-field image. (e) Respective binary images. (f) Decrypted volumetric data using digitized images in (e), showing a high-quality reconstruction comparable to Figure 2c using the full light-field information. Scale bar in (c), 4 μm.

**Figure 1**

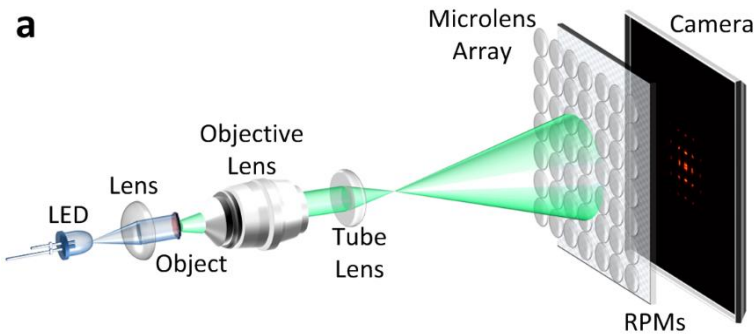

**Figure 2**

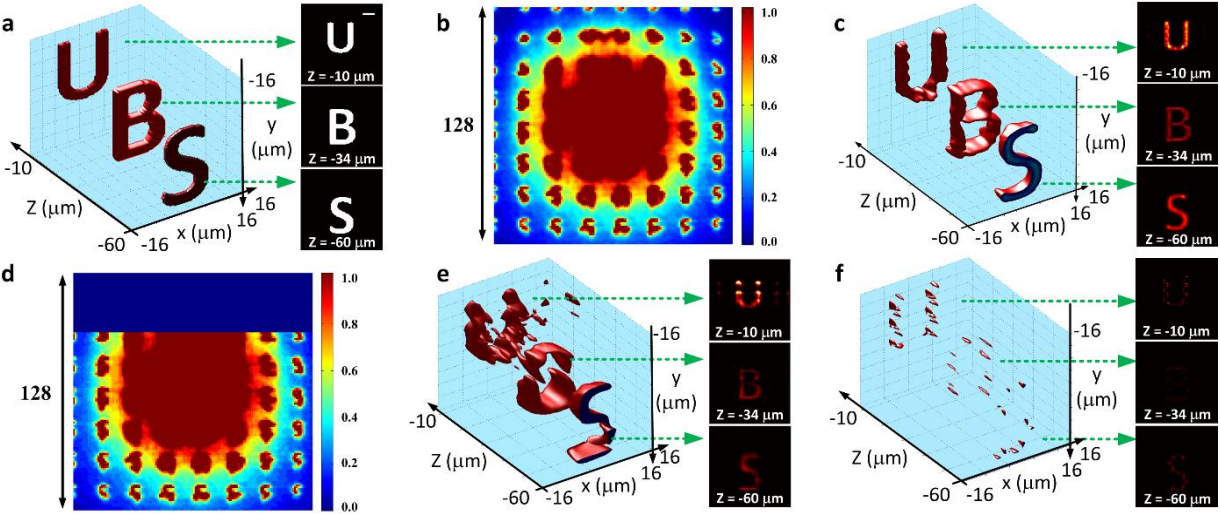

**Figure 3**

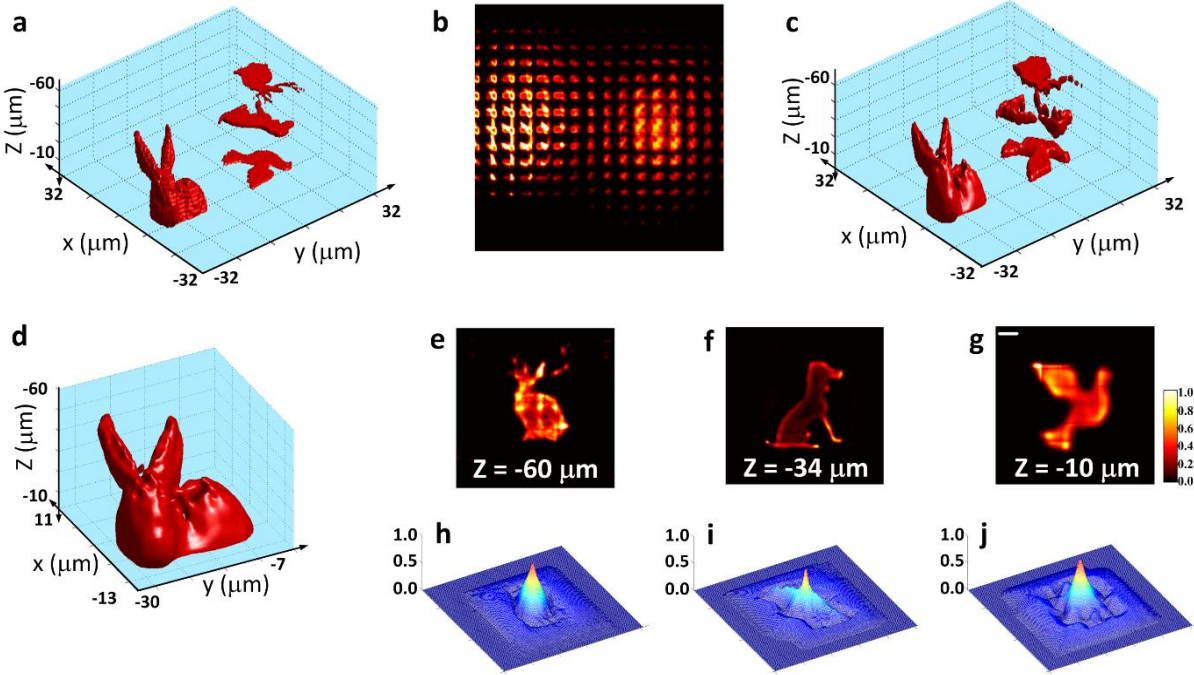

**Figure 4**

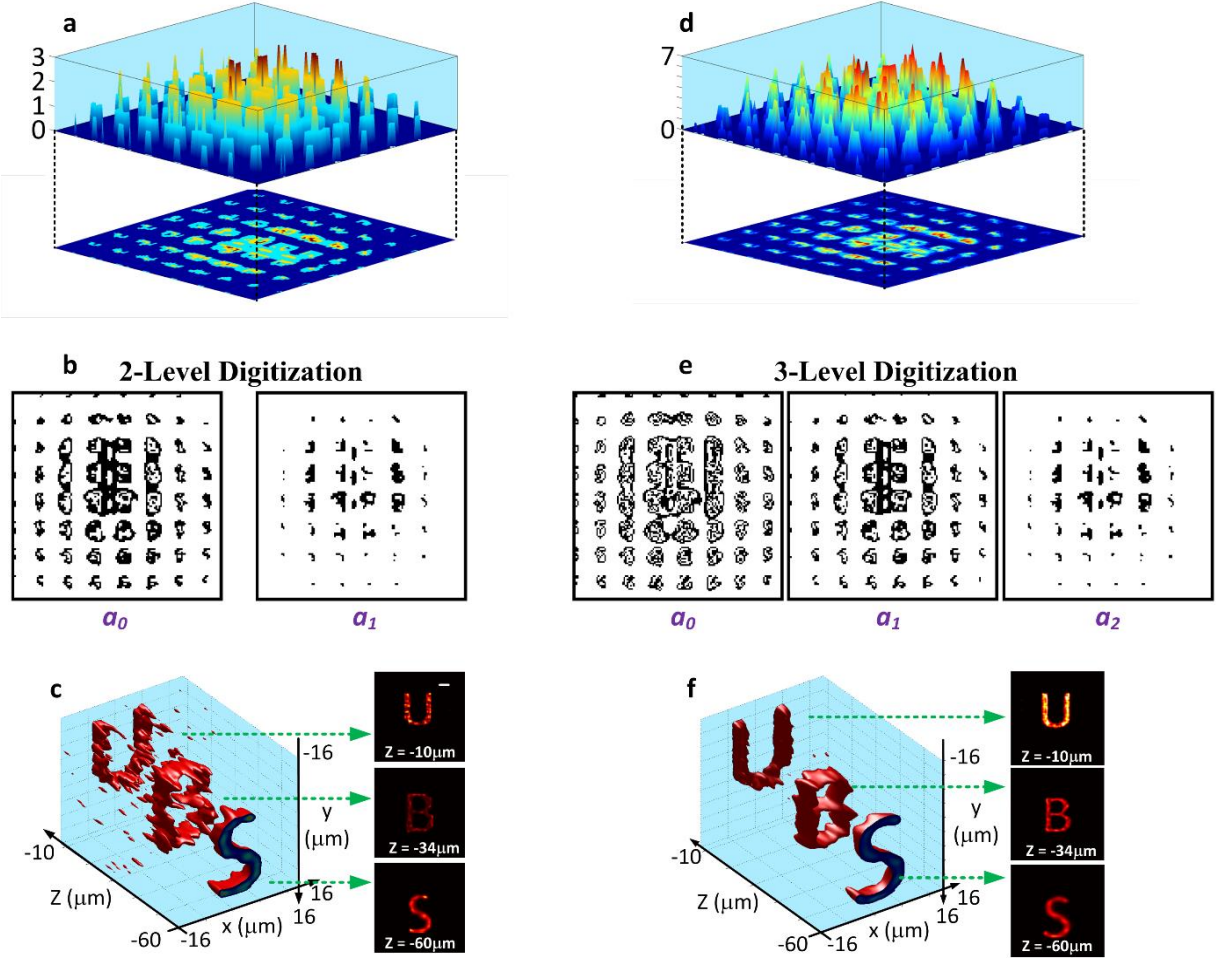

Supplementary Information for

# Volumetric Light-field Encryption at the Microscopic Scale

Haoyu Li, Changliang Guo, Inbarasan Muniraj, Bryce C. Schroeder, John T. Sheridan and Shu Jia

| | |
|---|---|
| Supplementary Figure 1 | Light-field encryption using random amplitude masks |
| Supplementary Figure 2 | Digitization for binary storage and read-out of the encrypted data using random amplitude masks |
| Supplementary Figure 3 | Volumetric encryption and decryption of non-binary objects |
| Supplementary Note | Pixel size of the numerical model of light-field propagation and image formation |

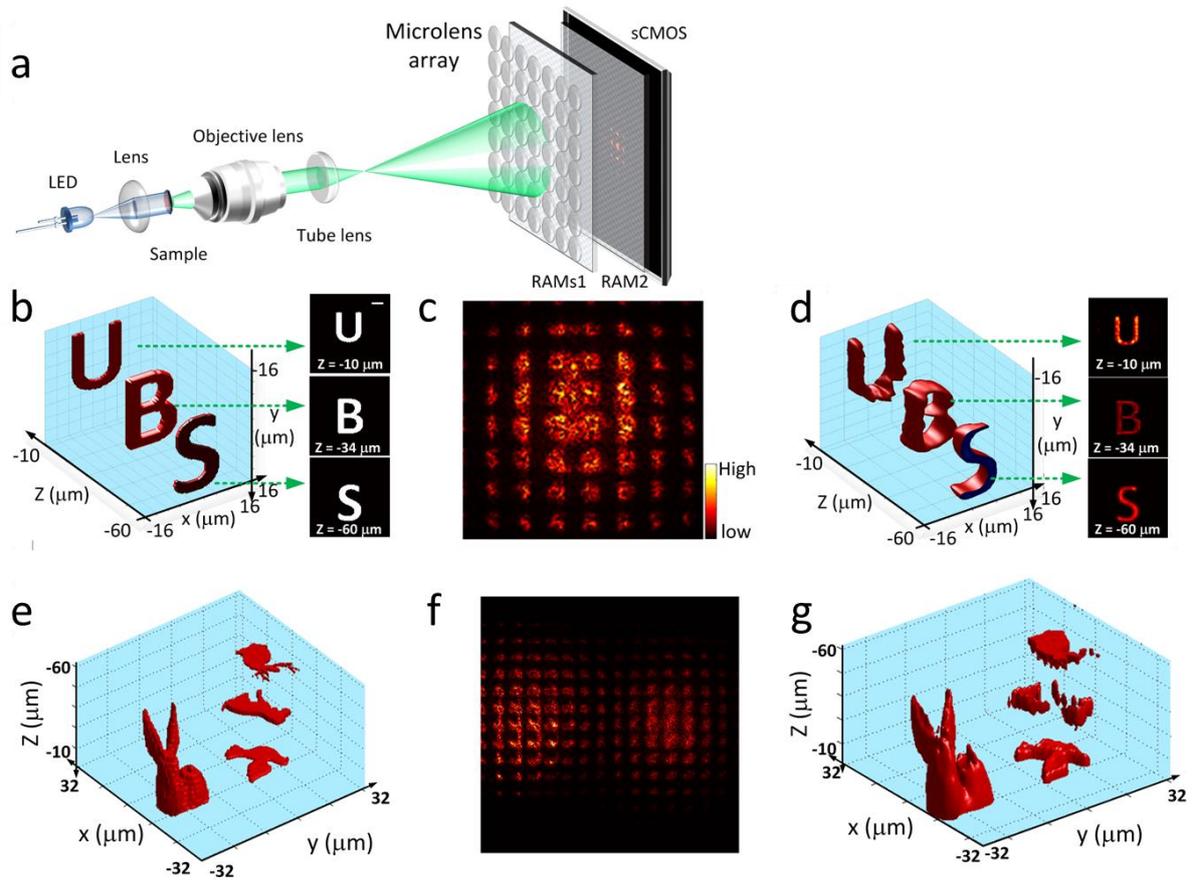

**Supplementary Figure 1. Light-field encryption using random amplitude masks (RAMs).** (a) Optical setup of the light-field encryption system. The object is imaged by the objective lens, and an intermediate image is formed by the tube lens. The microlens array is located at the intermediate image plane of the tube lens. An array of random amplitude masks (RAMs1) is located next to the microlens array. A camera sensor is placed at a distance of $f_{\mu lens}$ after the microlens array and RAMs1. The second random amplitude mask (RAM2) is located in front of the CCD camera. (b) Volumetric light-field encryption and decryption of three volumetric objects of the letters 'S', 'B' and 'U', located at -60 µm, -34 µm, and -10 µm along the axial (z) dimension, respectively. (c) The corresponding encrypted 128-pixel x 128-pixel 2D light-field image. Visual scrutiny has been largely suppressed. (d) Decrypted 3D volumetric information using the PSF key. (e) Optical encryption of multiplexed volumetric data, containing discrete and continuous volumetric data. (f) The encrypted 256-pixel x 256-pixel 2D light-field image. (g) Decryption and reconstruction of 3D volumetric information. Scale bar in (b), 4 µm.

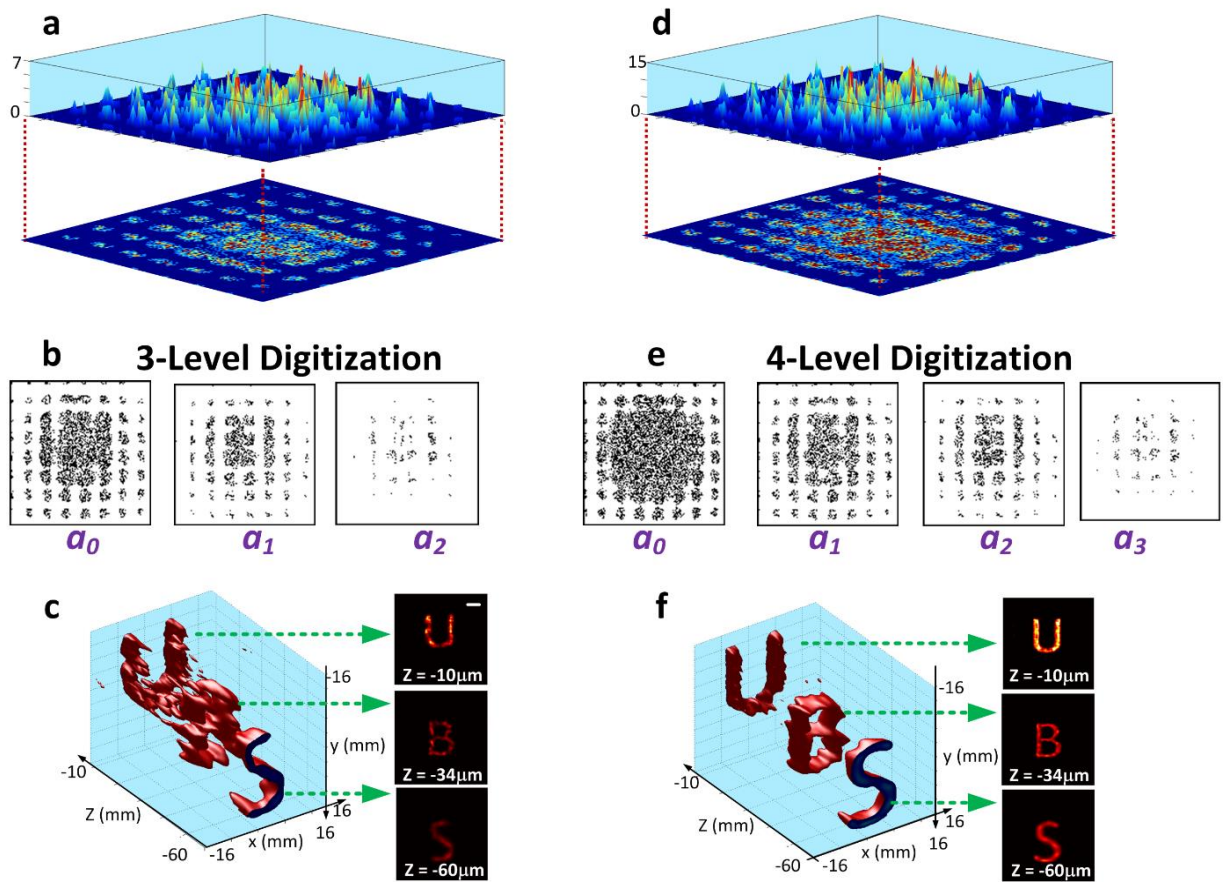

**Supplementary Figure 2. Digitization for binary storage and read-out of the encrypted data using random amplitude masks (RAMs).** (a) $N$ = 3 level digitization of the encrypted 2D light-field image in Figure 2b. Pixel values of the light-field image were normalized to $2^N - 1 = 7$ and rounded to the closest integer values $P$, which can be expressed as $P = a_0 \cdot 2^0 + a_1 \cdot 2^1 + a_2 \cdot 2^2$, where $a_i = 0 \ or \ 1 \ (i = 0, 1, 2)$ is the pixel-value composite coefficient. (b) The respective images that represent the binary composites of the pixel values of the light-field image at $2^0 = 1$, $2^1 = 2$, and $2^2 = 4$, respectively. (c) Decrypted volumetric data using digitized images in (b), revealing the original volumetric information. (d) $N$ = 4 level digitization of the same 2D light-field image. (e) Respective binary images. (f) Decrypted volumetric data using digitized images in (e), showing a high-quality reconstruction comparable to Supplementary Figure 1d using the full light-field information. Scale bar in (c), 4 µm.

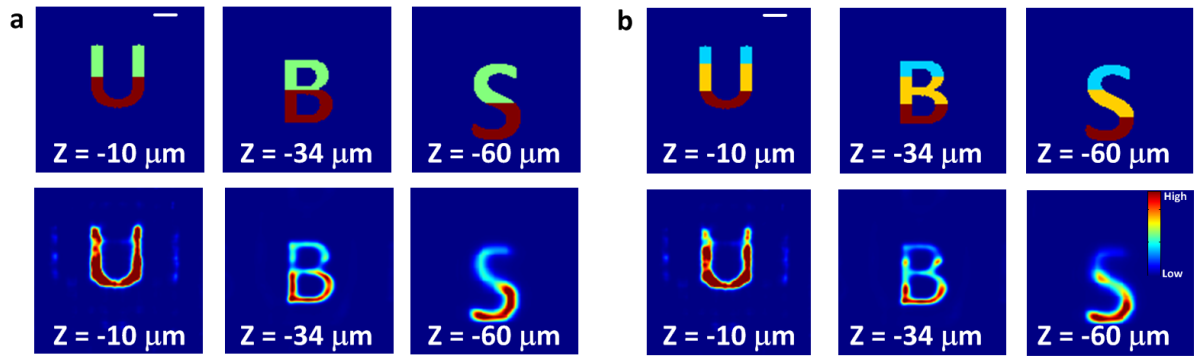

**Supplementary Figure 3. Volumetric encryption and decryption of non-binary objects.**
(a) Top row, 3D volumetric objects spatially organized as in Figure 2. The amplitude of the objects is distributed across three levels (0-blue, 128-green, 255-red). Bottom row, decrypted 3D information following the same procedures as in Figure 2. The variations in amplitude can be visualized. (b) Top row, the amplitude of the objects is distributed across four levels (0-blue, 85-light blue, 170-yellow, 255-red). Bottom row, decrypted 3D volumetric information. The variations in amplitude can be visualized. Scale bars, 4 µm.

**Supplementary Note -** Pixel size of the numerical model of light-field propagation and image formation

A *Spectral Method* [1,2] model is adopted in Eq. 5 in the manuscript (rewritten below as Supplementary Eq. 1), considering that the free space propagation in z is a near-field Fresnel patterns calculation. This process is given by performing the Fourier transform on the input information followed by multiplying the spectrum chirp, and finally the inverse Fourier transform is taken on the resulting information in the Fourier space.

$$h(\mathbf{x}, \mathbf{p}) = \mathcal{F}^{-1}\left\{\mathcal{F}[U_i(\mathbf{x}, \mathbf{p})\Phi(\mathbf{x})R(\mathbf{x})] \times \exp\left[i2\pi\lambda f_{\mu lens}\sqrt{1 - (f_x^2 + f_y^2)}\right]\right\} \quad (1)$$

It is worthwhile to note that, although both the *Spectral Method* (in near-field Fresnel pattern calculation) and the *Direct Method* (in far-field Fresnel pattern calculation), discussed in details in the references below, adopted on performing the Fresnel transform are mathematically equivalent, the oscillatory behavior of the kernels makes them computationally different[1].

Thus discretization of Supplementary Eq. 1 will provide different application conditions from those in far-field Fresnel patterns calculation. A sampled version of Supplementary Eq. 1 is:

$$h(\mathbf{x}', \mathbf{p}) = \mathcal{F}^{-1}\left\{\mathcal{F}[U_i(\mathbf{x}', \mathbf{p})\Phi(\mathbf{x}')R(\mathbf{x}')] \times \exp\left[i2\pi\lambda f_{\mu lens}\sqrt{1 - \left((m''\Delta f_x)^2 + (n''\Delta f_y)^2\right)}\right]\right\} \quad (2)$$

where $\mathbf{x}' = \{m'\Delta x, n'\Delta y\}$, $\Delta x$ and $\Delta y$ are the sampling intervals in x and y direction in the input and output space, and $m'$ and $n'$ are integer sampling indices in the input space. $\Delta f_x$ and $\Delta f_y$ are the sampling intervals in Fourier space, where $\Delta f_x = \Delta f_y = \frac{1}{N\Delta x}$, and the Fourier coordinates $f_x = m''\Delta f_x$ and $f_y = n''\Delta f_y$. *N* is the sampling number of the point spread function (PSF) in lateral plane (x or y). $m''$ and $n''$ are integer sampling indices in Fourier space.

Admitting sampling just in the Nyquist limit[2], the range of distances can be calculated. The application condition is near-field Fresnel propagation in our case. The Nyquist condition discussed in the reference paper[1] can be written as:

$$f_{\mu lens} \leq \frac{N(\Delta x)^2}{\lambda} \quad (3)$$

In our model, the sampling number of the PSF is N = 151, and Δ*x* is the sampling interval on the microlens array plane, i.e. the pixel size. The propagation distance z equals the focal length of the lenslet ($f_{\mu lens}$ = 3 mm). The wavelength is $\lambda$ = 532 nm in our case. Therefore, the condition for the sampling pixel size is given as

$$\Delta x \geq 3.25 \text{ μm} \quad (4)$$

In our case the sampling interval is chosen to be 10 μm which satisfies the sampling condition.